\documentstyle[color,epsfig,amssymb]{aipproc}

\begin{document}
\bibliographystyle{elsevier}
\title{A Compton Backscattering Polarimeter for Measuring Longitudinal Electron Polarization}

\author{\underline{I.  Passchier}$^\star$,
D.~W.~Higinbotham$^{\dagger}$,
N.~Vodinas$^{\star}$\thanks{Present address: IASA,
  P.O.Box 17214, 10024 Athens, Greece},
N.~Papadakis$^{\star 1}$, 
C.~W.~de Jager$^{\star}$\thanks{Present address: TJNAF, 
  Newport News, VA 23606, USA},
R.~Alarcon$^\ast$,
T.~Bauer$^\star$,
J.~F.~J.~van~den~Brand$^{\bullet,\star}$,
D.~Boersma$^\star$,
T.~Botto$^\star$,
M.~Bouwhuis$^\star$,
H.~J.~Bulten$^\bullet$,
L.~van~Buuren$^\bullet$,
R.~Ent$^\times$,
D.~Geurts$^\bullet$,
M.~Ferro-Luzzi$^{\star,\diamond}$,
M.~Harvey$^\times$,
P.~Heimberg$^\bullet$,
B.~Norum$^\dagger$,
H.~R.~Poolman$^\star$, 
M.~van~der~Putte$^\star$,
E.~Six$^\ast$,
J.~J.~M.~Steijger$^\star$,
D.~Szczerba$^\bullet$,
H.~de~Vries.$^\star$}
 
\address{
$^\star$NIKHEF, P.O.Box 41882, 1009 DB Amsterdam, Netherlands\\
$^{\dagger}$Department of Physics, University of Virginia,
			Charlottesville, VA 22901, USA\\ 
$^\ast$Department of Physics, Arizona State University, Tempe, AZ 85287, USA\\
$^\bullet$Department of Physics and Astronomy, Vrije Universiteit,
Amsterdam, The Netherlands\\
$^\times$TJNAF, Newport News, VA 23606, and Hampton University, Hampton, VA 23668, USA\\
$^\diamond$Institut f\"ur Teilchenphysik, Eidg. Technische Hochschule, CH-8093
Z\"urich, Switzerland\\
}
\maketitle

\begin{abstract}
Compton backscattering polarimetry provides a fast measurement of the
polarization of an electron beam in a storage ring.  Since the method
is non-destructive, the polarization of the electrons can be monitored
during internal target experiments.  At NIKHEF a Compton polarimeter
has been constructed to measure the polarization of the longitudinally
polarized electrons stored in the AmPS ring.  First results obtained
with the polarimeter, the first Compton polarimeter to measure the
polarization of a stored longitudinally polarized electron beam, are
presented in this paper.
\end{abstract}

\section*{Introduction}
 
The NIKHEF Compton polarimeter has been constructed to measure the
longitudinal polarization of electrons stored in the AmPS ring.  The
polarized electrons are provided by a recently commissioned polarized
electron source (PES)~\cite{cbp:bol96}.  While Compton
backscattering polarimeters are used to measure the polarization of
transversely polarized stored electron beams\cite{cbp:pla89,cbp:bar93},
NIKHEF's detector was the first to measure the polarization of a
longitudinally polarized stored beam\cite{cbp:igo96}. 
 
In this technique, a circularly polarized photon beam (polarization
$S_3$, energy $E_\lambda$) is backscattered from a stored polarized
electron beam (polarization $P_e$, energy $E_e$).

The cross section for Compton scattering of circularly polarized photons
from longitudinally polarized electrons can be written as
\begin{equation}
\frac{d\sigma}{dE_{\gamma}}=
\frac{d\sigma_{0}}{dE_{\gamma}}[1+S_{3}P_{z}\alpha_{3z}(E_{\gamma})],
\label{eq229}
\end{equation}
where $\frac{d\sigma_{0}}{dE_{\gamma}}$ follows from the energy
spectrum for unpolarized electrons and photons and $P_z$ represents
the longitudinal component of the electron polarization.  For a given
$E_{\lambda}$ and $E_{e}$ the asymmetry can be written as,
\begin{equation}
A(E_{\gamma})=\frac{N_L(E_{\gamma})-
     N_R(E_{\gamma})}{N_L(E_{\gamma})+N_R(E_{\gamma})}
={\Delta}S_3P_{z}\alpha_{3z}(E_{\gamma})
\label{eq:genasym}
\end{equation}
where $N_L(E_{\gamma})$ ($N_R(E_{\gamma})$) is the number of photons
with energy $E_{\gamma}$ with incident left (right) handed helicity,
and ${\Delta}S_3$ is the difference between the two polarization states,
divided by two.  $P_{e}$ is determined by taking $P_z$ as a free
parameter and fitting the measured asymmetry with
eq.~\ref{eq:genasym}. The relation between $P_z$ and $P_e$ is determined
by the lattice of the storage ring.

\section*{Layout of the polarimeter}
 
\begin{figure}
\centering{\epsfig{figure=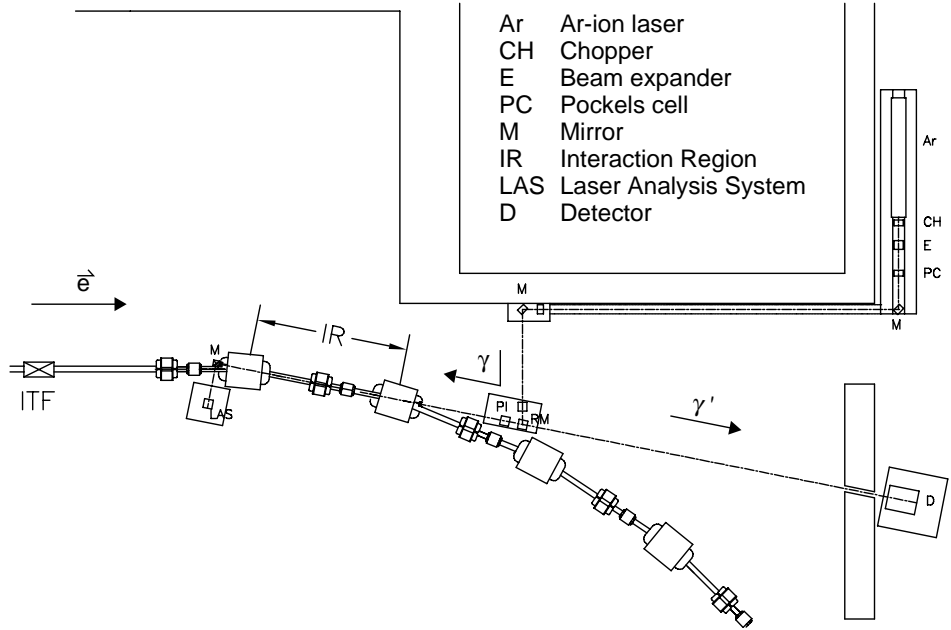,angle=-90,width=12.5cm}}
\caption{Schematic layout of the Compton polarimeter at NIKHEF. Indicated
is part of the AmPS ring, the optical system, the gamma-detector and the
internal target facility.}
\label{fig:layout}
\end{figure}

A schematic layout of the Compton polarimeter is shown in
fig.~\ref{fig:layout}.  The polarimeter consists of a laser system
with its associated optical system and a detector for the detection of
backscattered photons.  

Laser photons are produced by a 10~W CW Ar-ion laser, operated at
514~nm.  Part of the mirrors in the optical path can be controlled
remotely, in order to optimize the overlap of the electron and laser
beam.  A quarter-wave plate is used to convert the initially linearly
polarized photons to circularly polarized.  A Pockels cell is used to
switch the helicity between left and right, while a half-wave plate
can be inserted in the optical path to check for false asymmetries by
reversing the sign of the Compton asymmetry.

Laser photons interact with stored electrons in the straight section
(length $\approx 3$~m) between the first dipole and second dipole
(bending angles $11.25^\circ$) after the internal target facility. 
The backscattered photons leave the interaction region traveling in
the same direction as the electrons of the beam and are separated from
them after the second dipole.  They are detected in a gamma detector,
consisting of a block of $100\times100\times240$~mm$^3$ pure CsI.
 
A chopper mounted immediately after the Ar-ion laser is used to block
the laser light for $1/3$ of the time for background measurements. 
The chopper is operated at 75~Hz and also generates the driving signal
for the Pockels cell.

\section*{Results}
\label{sec:results}
 
The storage ring could only be operated with a 10\% partial
snake\cite{cbp:ohm96}.  Therefore, it was necessary to perform all
measurements with an electron beam energy of 440~MeV, resulting in a
maximum energy for the Compton photons of 7.04~MeV.  This energy is
lower than that of the design specification (500--900~MeV), resulting
in a poor energy resolution.  To reduce background at this rather low
energy, we performed all measurements with beam currents smaller than
15~mA.  The rate of backscattered photons was in the order of 8~kHz/mA
at full laser power, in agreement with simulations.

\begin{figure}
\centering{\epsfig{figure=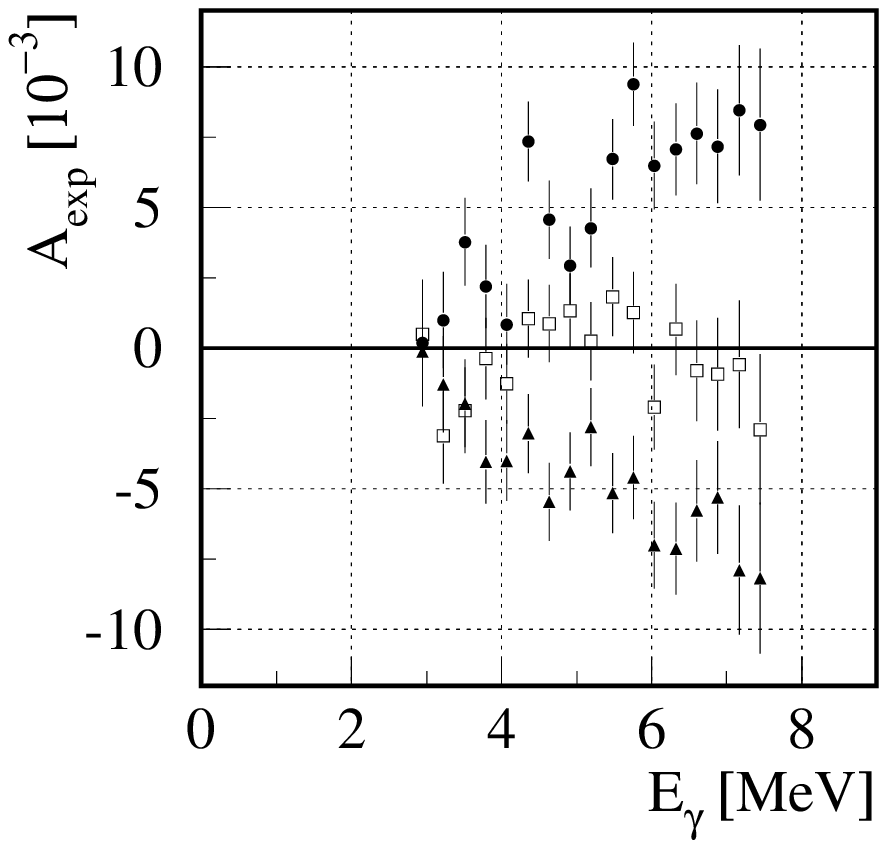, width=6.7cm}
\epsfig{figure=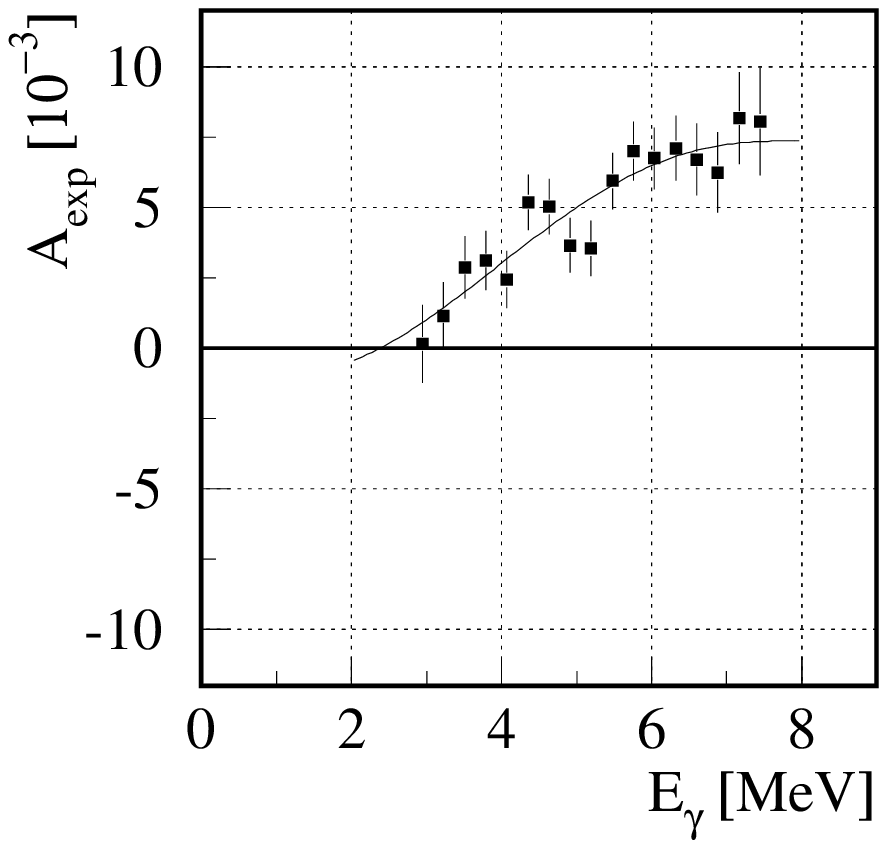, width=6.7cm}}
\caption{Left: Raw asymmetries for electrons with
left~($\bullet$) and right-handed helicity~($\blacktriangle$) and for
unpolarized electrons~($\square$). Right: the average of the
asymmetries for electrons with left and right-handed helicity, taking into
account the difference in sign of the asymmetry, after correcting for
false asymmetries.}
\label{fig:asymspec}
\end{figure}

To minimize the effects of false asymmetries (induced by a small
steering effect of the Pockels Cell), we performed sets of six
independent measurements to determine the electron polarization. 
Three measurements were done with different electron polarizations
injected into the ring (positive helicity, unpolarized and negative
helicity).  These measurements were repeated with the half-wave plate of
the polarimeter inserted in the optical path.  The measurements with
unpolarized electrons were used to determine and correct for false
asymmetries, while the insertion of the half-wave plate was done as a
consistency check. 
Figure~\ref{fig:asymspec} shows the asymmetry before and after
correction for false asymmetries.

\begin{figure}
\centering{\epsfig{figure=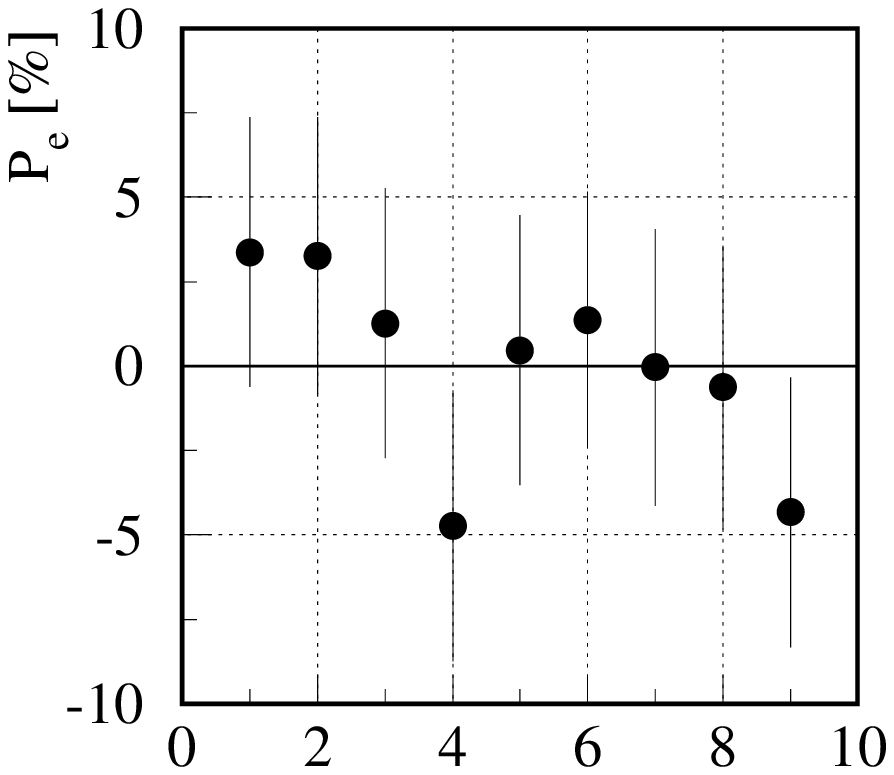,width=6.7cm}
\epsfig{figure=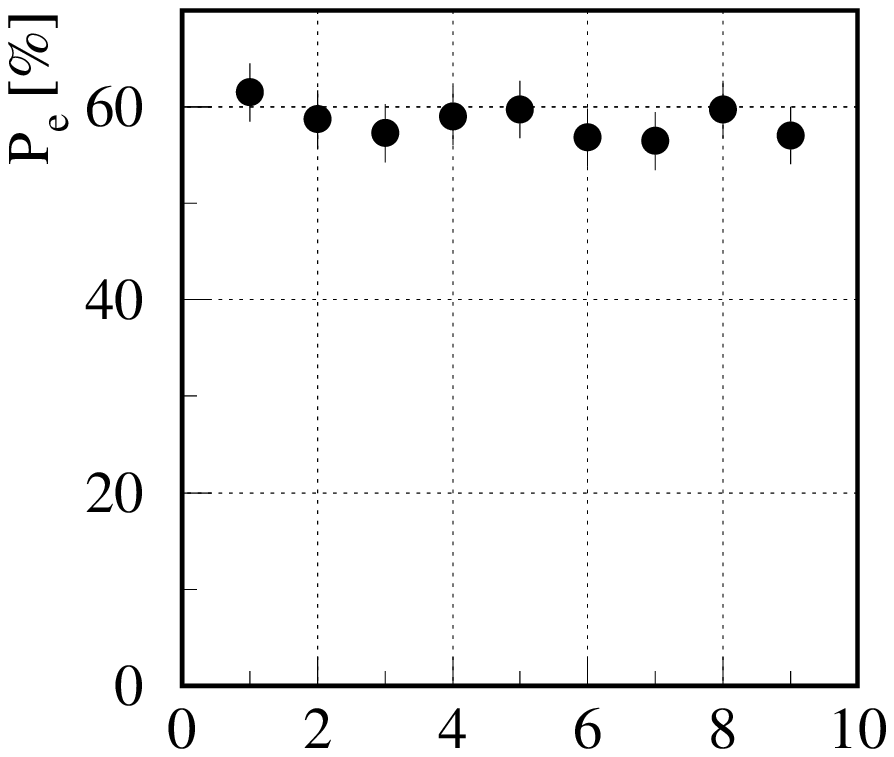,width=6.7cm}}
\caption{Left: short-term stability of the polarimeter. Every data point
represent a measurement of the polarization of unpolarized electrons. The
time between two measurements is $\approx$10~min.
Right: long-term stability of the polarimeter. Every data point
represents a complete set of six polarization measurements. The interval
between the measurements is typically one day.}
\label{fig:reprolong}
\label{fig:reproshort}
\end{figure}

To determine the stability of the polarimeter, one measurement was
repeated nine times.  To exclude any sensitivity to variations in the
polarization of the injected electrons or spin life time, those
measurements were performed with unpolarized electrons.  The total
measurement time was $\approx$~90~min, while a full set of six
measurements normally takes about 60~min.  The results are shown in
fig.~\ref{fig:reproshort} and show good stability on this time
scale.

The long-term stability is determined from polarization measurements
done typically once a day.  These measurements are sensitive not only
to variations of the polarimeter, but also to any other time-dependent effect
such as a degradation of the
cathode used at PES.
The results (see fig.~\ref{fig:reprolong}) show no trend in the
polarization of the electrons, indicating a good long-term stability
for all components.

The polarimeter has been used successfully to optimize the settings of
the Z-manipulator at PES.  After the optimization, the spin life
time~($\tau$) and initial polarization~($P_0$) has been determined by
combining the data of nine measurements of six minutes each.  The
combined data have been rebinned as a function of time and the
polarization has been determined for each bin separately.
We found $P_0 = 61.6 \pm 1.4 \text{\% (statistical)}$,
and $\tau = 4500^{+5900}_{-1600}\text{~s}$.  The spin life time is in
agreement with our calculations.  The polarization measured with the
Mott polarimeter at PES was $82 \pm 5 \%$.  The difference between the
polarization measured by the Mott polarimeter and by the Compton
polarimeter may be caused by depolarization due to the
focusing solenoids in the linac or depolarizing resonances during
damping of the beam.

\section*{Conclusions} 

Here, we describe the results of extensive tests done with a Compton
backscattering electron polarimeter.  The tests have
been performed at an electron energy of 440~MeV and a partial snake.  The
results show that it is possible to operate the polarimeter in a
reliably manner over a period of weeks.  Furthermore, the polarimeter
has been used to map out the full dependence of the electron
polarization of stored electrons on the settings of the Z-manipulator, and
to determine the spin life time and depolarization during acceleration
and injection of the electrons.

\section*{Acknowledgment}

This work was supported in part by the Stichting voor Fundamenteel Onderzoek
der Materie (FOM), which is financially supported by the Nederlandse
Organisatie voor Wetenschappelijk Onderzoek (NWO), the Swiss
National Foundation, the National Science
Foundation under Grants No. PHY-9316221 (Wisconsin), PHY-9200435 (Arizona
State) and HRD-9154080 (Hampton), Nato Grant No. CRG920219.
and HCM Grant Nrs.  ERBCHBICT-930606 and ERB4001GT931472.

\enlargethispage{1cm}
\bibliography{polarimeter}

\begin{thebibliography}{1}

\bibitem{cbp:bol96}
Y.~B. Bolkhovityanov {\em et~al.},
\newblock The polarized electon source at {NIKHEF},
\newblock in {\em Proc. of the $12^{th}$ International Symposium on High Energy
  Spin Physics}, edited by C.~W. de~Jager {\em et~al.}, pages 730--732, World
  Scientific, 1996.

\bibitem{cbp:pla89}
M.~Placidi and R.~Rossmanith,
\newblock Nucl. Instr. Meth. Phys. Res. {\bf A274} (1989) 79.

\bibitem{cbp:bar93}
D.~P. Barber {\em et~al.},
\newblock Nucl. Instr. Meth. Phys. Res. {\bf A329} (1993) 79.

\bibitem{cbp:igo96}
I.~Passchier {\em et~al.},
\newblock A {C}ompton backscattering polarimeter for electron beams below 1
  {GeV},
\newblock in {\em Proc. of the $12^{th}$ International Symposium on High-Energy
  Spin Physics}, edited by C.~W. de~Jager {\em et~al.}, pages 807--809, World
  Scientific, 1996.

\bibitem{cbp:ohm96}
C.~Ohmori {\em et~al.},
\newblock Phys. Rev. Lett. {\bf 76} (1996).

\end{thebibliography}






\end{document}